\documentclass[english,aps,prstper,reprint,showpacs]{revtex4-1}
\usepackage[T1]{fontenc}	
\usepackage[latin9]{inputenc}	
\usepackage{geometry}
\usepackage{graphicx}
\usepackage[above,below]{placeins}	
\usepackage{times}
\setcitestyle{numbers,square}	
\begin{document}

\title{Community structure in introductory physics course networks}
\author{Adrienne L. Traxler}
\affiliation{Wright State University, Department of Physics,\\ 3640 Colonel Glenn Highway, Dayton, OH 45435}


\begin{abstract}
Student-to-student interactions are foundational to many active learning environments, but are most often studied using qualitative methods. Network analysis tools provide a quantitative complement to this picture, allowing researchers to describe the social interactions of whole classrooms as systems. Past results from introductory physics courses have suggested a sharp division in the formation of social structure between large lecture sections and small studio classroom environments. Extending those results, this study focuses on calculus-based introductory physics courses at a large public university with a heavily commuter and nontraditional student population. Community detection network methods are used to characterize pre- and post-course collaborative structure in several sections, and differences are considered between small and large classes. These results are compared with expectations from earlier findings, and comment on implications for instruction and further study.
\end{abstract}


\maketitle

\section{Introduction}

Social interactions are a fundamental part of most classroom environments,
especially those with an interactive engagement (IE) focus. Despite its importance, 
the structure of student classroom community is not typically examined in detail or 
in conjunction with other outcome measures such as
conceptual or attitudinal gains. One resource to enrich this perspective comes from 
the social sciences, where social network analysis has been
used for decades  as a way to quantify and explore
communities and the interactions that structure them~\citep{borgatti_network_2009}. Recently, discipline-based education researchers have taken up these tools to begin systematically mapping peer interactions and how they correlate with other indicators~\citep{brewe_changing_2010,goertzen_expanded_2013,bruun_talking_2013,grunspan_understanding_2014}. 
These investigations can reach a wider pool of students than in-depth interviews, and the resulting large-scale picture provides a valuable counterpart to fine-grained qualitative data~\citep{dawson_study_2008}. Expanding our understanding of student classroom community also links to research on retention and persistence, which highlights the importance of learning-related social ties~\citep{tinto_classrooms_1997}.

Network analysis tools have proven useful at different levels of resolution. For a temporally-evolving, detailed picture, \citet{bruun_talking_2013} analyzed weekly surveys asking students who they interacted with in a number of contexts (e.g., problem-solving, in-class socializing). By aggregating this information over the semester, a complex weighted picture of student interactions emerges. Other studies take a less sampling-intensive approach and administer only a few surveys during the semester~\citep{brewe_changing_2010,grunspan_understanding_2014} to capture large-scale patterns and changes in students' collaborative behavior. The work reported here takes the second approach, using pre- and post-course surveys to form a baseline picture of students' physics learning interactions.

In an earlier study comparing two different course formats, \citet{brewe_changing_2010} found divergent types of community structure depending on course type. Students in a larger traditional lecture class ($N=80$) began and ended the semester collaborating with few or no other students. In a smaller studio-format section ($N=30$), students also began with few connections, but by the end of the semester had formed a tightly connected class-wide network. 
These results suggest that it is possible to characterize complex relations of
student interaction through low-impact survey instruments. Further, Brewe and collaborators' work indicates that
formation of student community is one substantial and measurable distinguishing feature 
between different learning environments.

This paper presents the first phase of analysis for social network data collected from four sections of first-semester introductory physics courses. 
Future stages of the project will combine network data 
with attitudinal and conceptual surveys to explore interactions between shifts in those measures and the classroom learning community~\citep{brewe_extending_2013}.
To add detail to the picture provided by other researchers~\citep{brewe_investigating_2012,brewe_changing_2010}, I use network methods of community detection to characterize student collaborative structure. 
The sections below outline basic concepts for network analysis and for quantifying community in networks, compare results from the four sections surveyed, and contrast with expectations from previously published results.

\section{Methods}
\subsection{Context and data collection}
Wright State University is a large public research university (13,614 students in fall 2014). The student body is predominantly in-state, with 9\% transfer students (largely from area community colleges), 6\% veterans, 12\% African Americans, and 3\% Hispanic/Latino students. Most students commute, and many are employed off-campus, leading to community-building challenges similar to those raised in research on retention and persistence~\citep{tinto_classrooms_1997,gilardi_university_2011}.

The present study follows the data collection of \citet{brewe_changing_2010}. In the first and last weeks of the 15-week semester, students completed an electronic survey asking, ``Who do you work with to learn physics in this class?'' They selected names from the roster, and were additionally asked to name any collaborators missing from the list. This additional question captured students who added the class during the first week, and also permits later investigation of cross-section collaboration. 

The four sections surveyed represent a fairly typical sample of calculus-based physics I at the institution. They are taken from fall (section A) and spring (sections B--D) semesters. Sections A and B were large lecture sections with a small to moderate degree of interactive engagement (some Peer Instruction~\citep{mazur_peer_1997} in lecture) and a weekly problem-solving recitation. 
Sections C and D were smaller ($N\leq 30$) sections with lecture and recitation combined, with a high degree of interactive engagement. Students in all sections took the same one-credit traditional laboratory course. Table \ref{tab:sections} summarizes information about the surveyed classes.

\begin{table}[htbp]
\caption{Enrollment and survey response rates for all sections.\label{tab:sections}}
\begin{tabular}{c@{\hskip 15pt}cc@{\hskip 15pt}cc}
\hline \hline
 & \multicolumn{2}{c@{\hskip 15pt}}{\textbf{Pre}} & \multicolumn{2}{c}{\textbf{Post}} \\ 
Section & Enrolled & Response & Enrolled & Response \\ \hline
A & 215 & 89\% & 209 & 63\% \\
B & 192 & 83\% & 188 & 81\% \\
C & 26 & 65\% & 19 & 84\% \\
D & 28 & 75\% & 26 & 69\% \\
\hline \hline
\end{tabular}
\end{table}

\subsection{Network analysis}
Collected survey data was cast in the form of a network object, which is comprised of nodes and links between nodes, or ``edges''. Each node represents a student, and an edge between two nodes indicates that one named the other as a collaborator in learning physics. Links are treated as undirected---i.e., an edge is present if either party indicates it, without considering who named who. 

Two basic descriptors of a network are its number of nodes and number of edges. To compare the relative interconnectedness of a section at the beginning and end of the semester, the network density $D$ is defined as the fraction of possible edges that exist. 
In undirected networks, the maximum number of edges is $N(N-1)/2$, where $N$ is the number of nodes.
A larger network has many more possible relationships, so will typically be lower density, thus it is not necessarily meaningful to compare density between large and small sections. However, within a single class, we would expect to see an increase in density if students begin to work together more freely during the semester. 

Network statistics inherently violate the assumption of independent measurements, so bootstrap methods are needed for any standard error and $t$-test calculations. Following the method detailed in \citet{snijders_non-parametric_1999}, 1000-sample bootstrap calculations were used to estimate standard errors for network densities. To look for paired pre-post differences in each section, networks were reduced to matched nodes and a $t$-test comparison was run on the pre- and post-densities.

Communities---groups of students who are significantly more linked with each other than with surrounding nodes---provide a more detailed way to measure the level of student collaboration in learning. In curricula where transforming participation or building connections among the student cohort are instructional goals~\citep{brewe_changing_2010,roychoudhury_gender-inclusive_1995}, community-finding metrics estimate the success of such efforts.
Detection of community structure is an active area of research in network analysis with a variety of possible methods \citep{fortunato_community_2010}. 
Here, community partitioning is performed using the Infomap algorithm~\citep{rosvall_maps_2008}, which uses an information theory perspective to highlight groups sharing a substantially larger information flow among each other than with their neighbors. In other words, any given piece of information (such as insight on how to approach a homework problem) is more likely to circulate among members of a community than to cross to outside students. 
Isolated students with no collaborators are each detected as individual ``communities.'' Thus, a decrease in the number of communities over the semester indicates that fewer students are working in isolation.

\section{Results}
\begin{figure}
\includegraphics[width=0.92\linewidth]{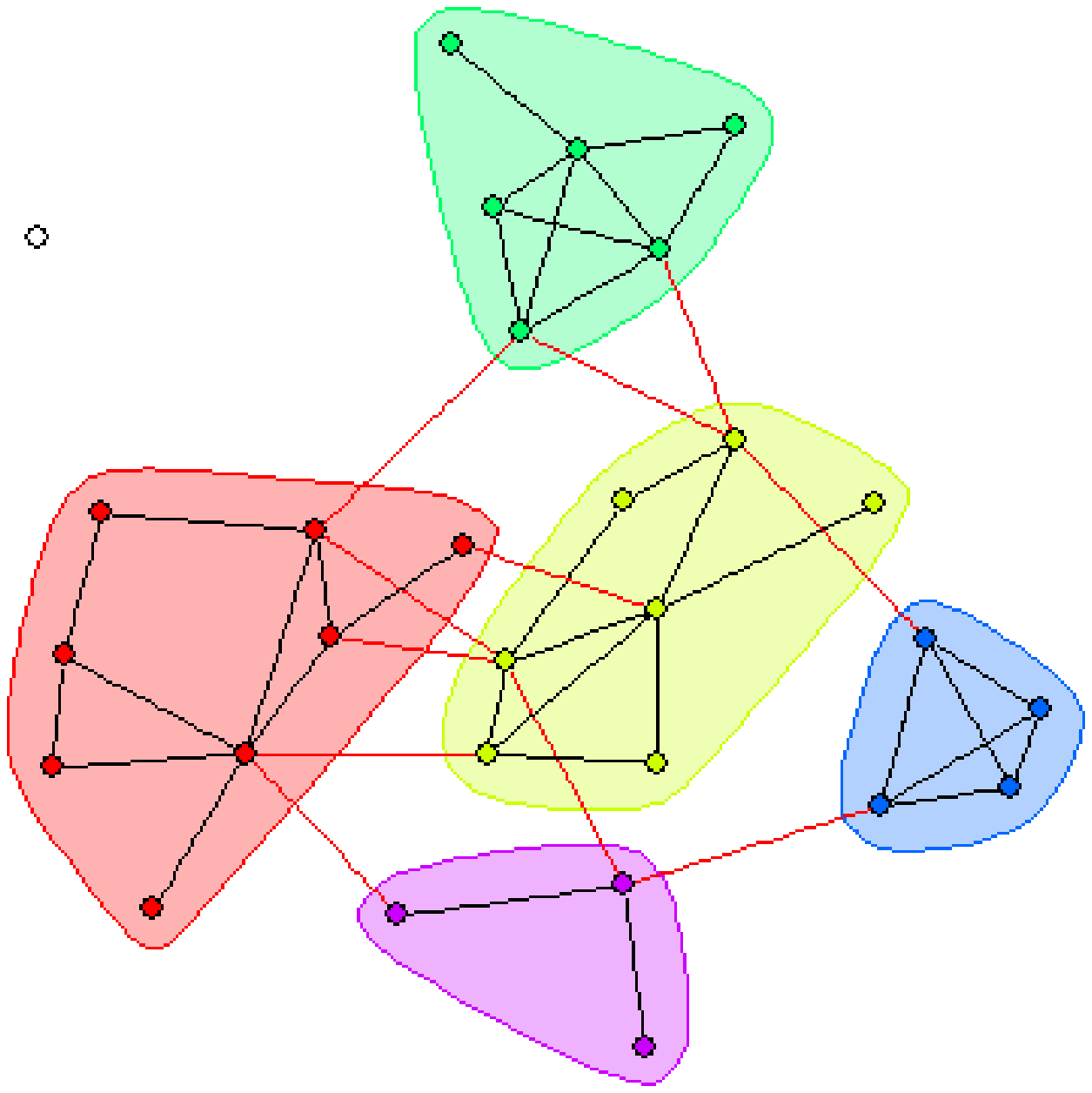}
\includegraphics[width=0.92\linewidth]{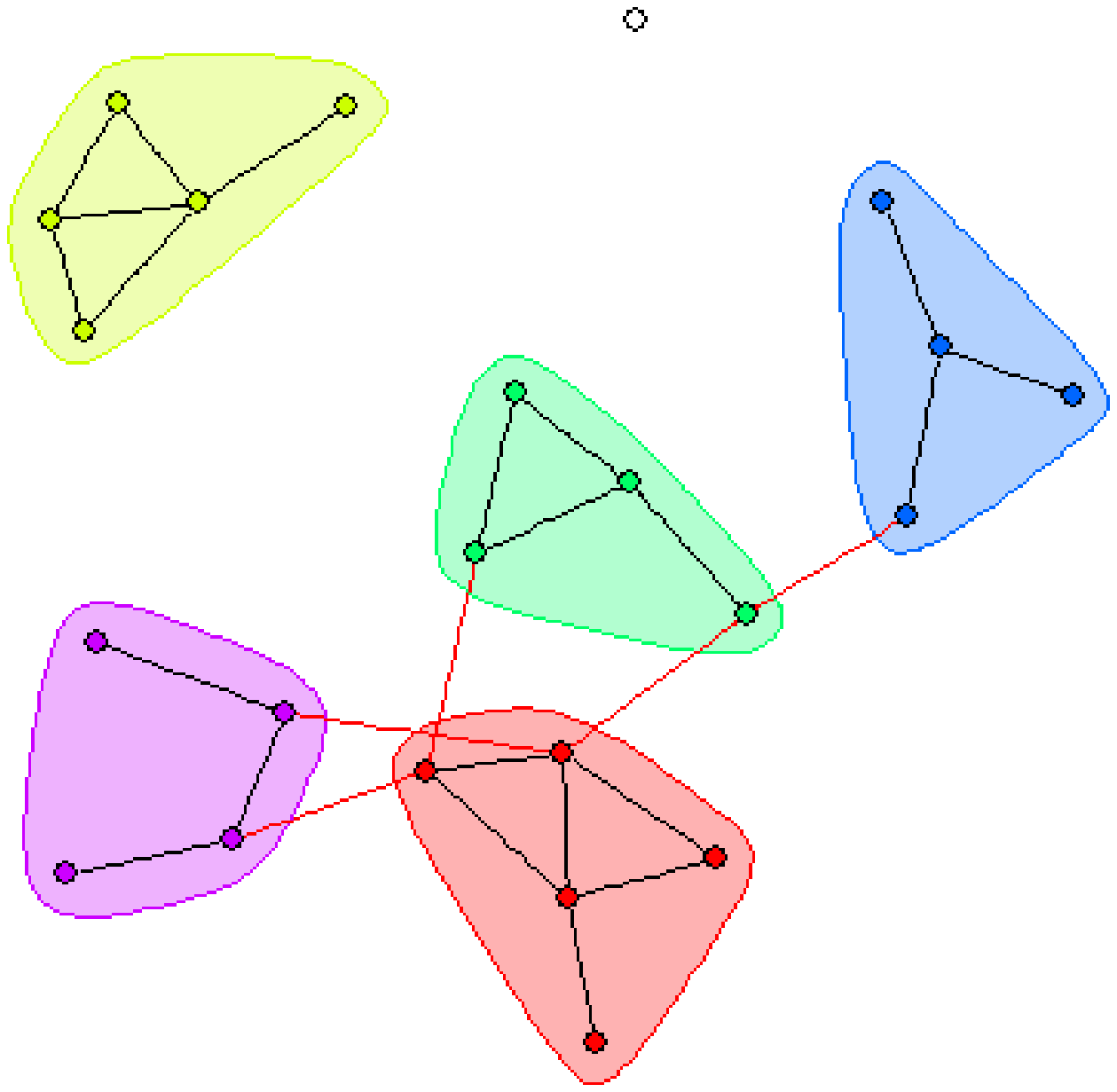}
\caption{Sociogram of pre-course (top) and post-course (bottom) collaborative networks for section D. Nodes and surrounding highlight cells are colored by community cluster. Isolated nodes add to the community count in Table \ref{tabdesc}, but are not colored on the diagram.\label{socio03J}}
\end{figure}

\begin{figure}
\includegraphics[width=0.92\linewidth]{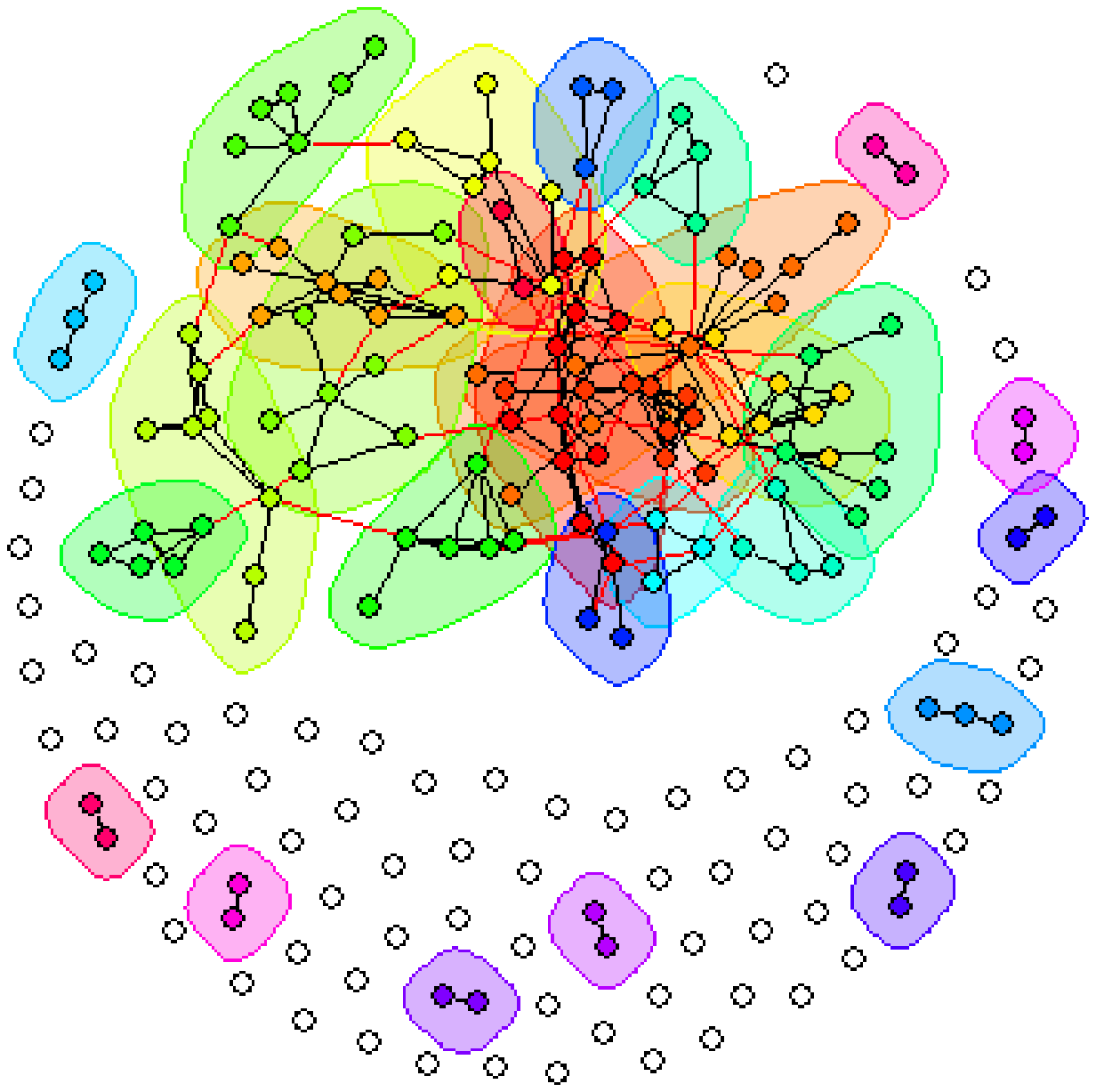}
\includegraphics[width=0.92\linewidth]{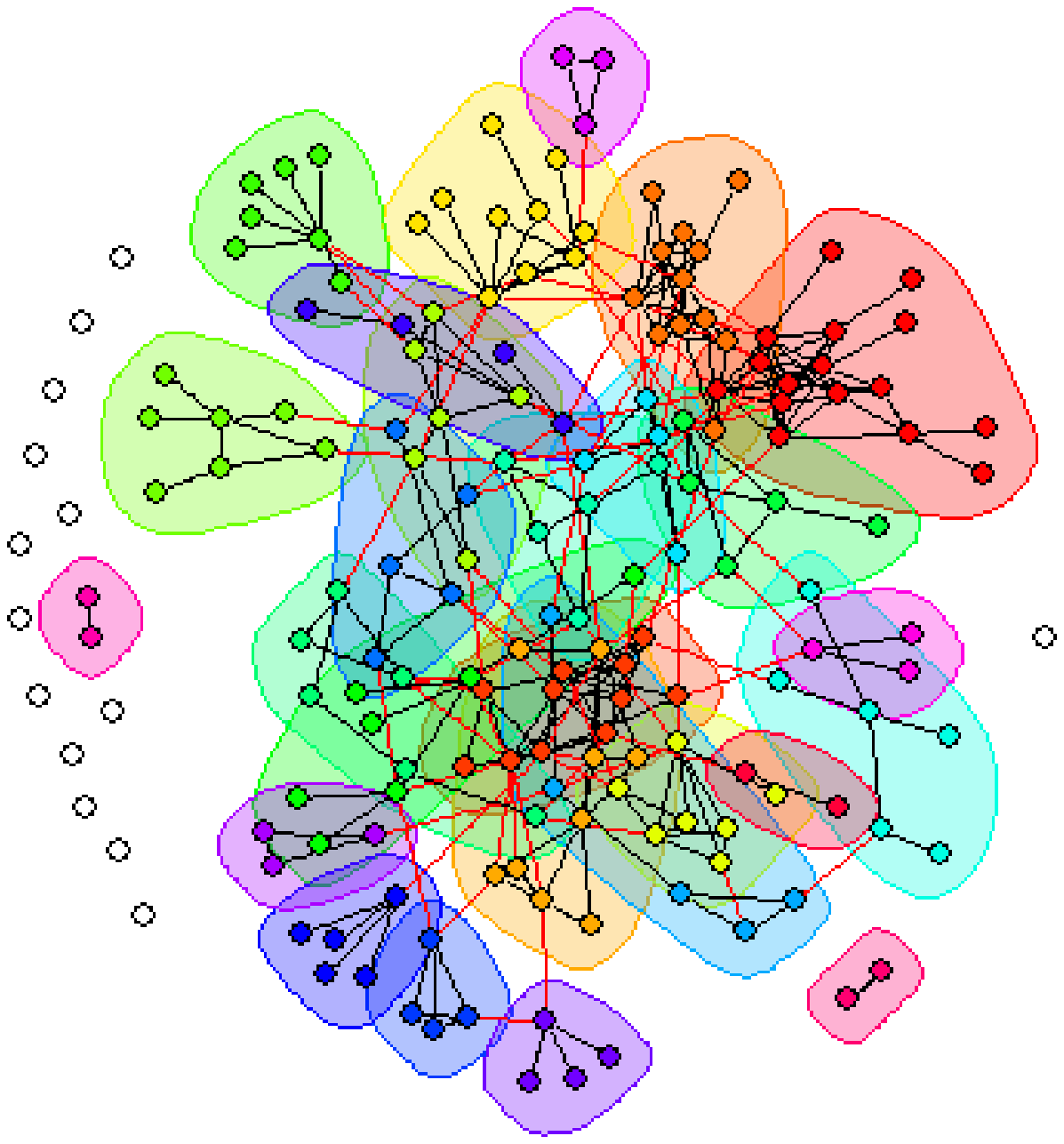}
\caption{Sociogram of pre-course (top) and post-course (bottom) collaborative networks for section A. Even in the top diagram, there is a large component of connected students. Red lines indicate links bridging communities.\label{socio01}}
\end{figure}

Table \ref{tabdesc} gives the basic descriptive statistics and calculated values for the networks from each of the four sections. These include number of nodes and edges, network density with bootstrap estimates of standard error (SE), bootstrap $t$-tests for pre-post density changes, and the number of communities detected by Infomap. Figures \ref{socio03J} and \ref{socio01} show the first- and last-week sociograms (diagrams of the network) for example small and large sections. Detected communities are used to group the nodes by color. 

\begin{table*}[htbp]
\caption{Characteristics of classroom networks. The number of nodes is equal to the number of survey respondents plus the named non-respondents, so it can exceed the participation rate listed on Table \ref{tab:sections}. Also included are the network density with standard errors (SE), and the total number of detected communities (\# Comm.). Finally, $t$-statistics and $p$-values are given for the pre-post change in density $\Delta D$.\label{tabdesc}}
\begin{tabular}{c@{\hskip 15pt}cccc@{\hskip 15pt}cccc@{\hskip 15pt}cc}
\hline \hline
 & \multicolumn{4}{c@{\hskip 15pt}}{\textbf{Pre}} & \multicolumn{4}{c}{\textbf{Post}} & \multicolumn{2}{c}{\textbf{Pre-Post $\Delta D$}} \\ 
Section & Nodes & Edges & Density (SE) & \# Comm. & Nodes & Edges & Density (SE) & \# Comm. & $t$ & $p$-value \\ \hline
A & 203 & 213 & 0.010 (0.002) & 97 & 174 & 304 & 0.020 (0.003) & 41 & 2.82 & 0.002 \\
B & 185 & 288 & 0.017 (0.031) & 50 & 177 & 327 & 0.021 (0.027) & 42 & 0.79 & 0.22 \\
C & 24 & 28 & 0.101 (0.003) & 7 & 19 & 17 & 0.099 (0.003) & 8 & -0.45 & 0.67 \\
D & 29 & 47 & 0.116 (0.038) & 6 & 23 & 27 & 0.107 (0.030) & 6 & -1.01 & 0.84 \\
\hline \hline
\end{tabular}
\end{table*}

All sections show an interesting contrast to the results of \citet{brewe_changing_2010}: substantial pre-course connections exist even among students in large lecture sections, and even in small interactive sections, the density of the final network is not significantly changed from the first week of classes. From Table \ref{tabdesc}, only section A showed a statistically significant shift in network density, with the number of connections nearly doubling over the semester. This section also showed the largest change in number of detected communities, more than halving the initial number of groupings. In the smaller sections, on the other hand, both density and number of communities stayed effectively constant.

\section{Discussion}
The findings of \citet{brewe_changing_2010} seem to suggest that small interactive physics classes foster student community, while larger lecture classes do not. This tendency seems intuitive given the divergence in classroom dynamics and (often) instructor goals in those environments. Despite expectations of similar conditions in this data, these results show different community patterns emerging. One possible explanation for the presence of structure in the lecture classes (and its relative stagnation in the smaller sections) is differences in student population by institution. All sections surveyed in this paper were drawn from the calculus-based introductory sequence, where most of the students are engineering majors. There is also a large fraction of international students in the sample, many of whom share a country of origin and sometimes ties of friendship or family. Finally, the course is typically not taken in a student's first semester because of prerequisites. Taken together, these factors may cultivate a larger than typical degree of community among students in this sample before the semester even begins.

For instructors interested in building connections among students~\citep{tinto_classrooms_1997,dawson_study_2008}, these results pose an interesting dilemma. Much of the advice toward fostering group dynamics in IE classes assumes that most students enter the course isolated, with some small fraction of existing friends or study groups. Faculty are thus advised to separate friends where possible, so that students build new connections instead of holding to existing social dynamics that exclude other group members~\citep{michaelsen_team-based_2004}. However, in a class with a large number of pre-existing connections, this may not be possible (even if desirable). Students who are not part of this existing community, especially those who live off-campus and work many hours, may experience difficulty in ``breaking in.'' On the other hand, a large initial density suggests that some of the work of connecting students has already been done, so this interconnectivity may be leveraged if instructors can adapt to work with it productively.

Another issue to consider is  incomplete data. From Table \ref{tab:sections}, survey response rates were lower in most sections at the end of the semester. This effect is mitigated somewhat by the use of undirected links: students who did not take a survey, but were named by any other student, would still appear in the final network. Future data collection will add follow-up with instructors to boost response rates.

The goal of this early stage anaysis was to chart out network structure in several sample sections of introductory physics at the author's institution. The results diverged unexpectedly from previous work in seemingly similar circumstances~\citep{brewe_changing_2010} and suggest a different focus for local instructional and departmental efforts toward building community among students. Later analysis will explore measures of students' position and influence in the network and how these may relate to other measures of success such as conceptual~\citep{bruun_talking_2013} or attitudinal gains. Based on these preliminary results, additional measures such as Rovai's Classroom Community Scale~\citep{rovai_development_2002} and in-depth qualitative study~\citep{goertzen_expanded_2013} may shed further light on the different forms of student participation and collaboration in the course.

\bibliographystyle{apsrev}  
\bibliography{networks_PERC}  

\end{document}